\documentclass{PoS}

\newcommand{\preprintline}{\newline
\vskip -6.2cm
\rightline{\parbox{4cm}{\large\rm  DESY 11-199}}
\vspace{4.2cm}
}

\usepackage{subfigure}
\usepackage{amsmath,amssymb}

\title{Anomalous scaling in the random-force-driven Burgers equation: A Monte Carlo study \preprintline}

\ShortTitle{Anomalous scaling in the random-force-driven Burgers equation}

\author{\speaker{David Mesterh\'azy}\,\thanks{Current address: Insitut f\"ur Theoretische Physik, Philosophenweg 16, 69120 Heidelberg, Germany} \\
        Institut f\"ur Kernphysik, TU Darmstadt, Schlossgartenstrasse 9/2, 64289 Darmstadt, Germany\\
        E-mail: \email{mesterh@crunch.ikp.physik.tu-darmstadt.de}}

\author{Karl Jansen\\
        NIC, DESY Zeuthen, Platanenallee 6, 15738 Zeuthen, Germany \\
        E-mail: \email{Karl.Jansen@desy.de}}

\abstract{We present a new approach to determine the small-scale statistical behavior of hydrodynamic turbulence by means of lattice simulations. 
Using the functional integral representation of the random-force-driven Burgers equation we show that high-order moments of velocity differences satisfy anomalous scaling.
The general applicability of Monte Carlo methods provides the opportunity to study also other systems of interest within this framework.}

\FullConference{The XXIX International Symposium on Lattice Field Theory - Lattice 2011\\
July 10-16, 2011\\
Squaw Valley, Lake Tahoe, California}

\begin{document}

\section{Introduction}

The incompressible Navier-Stokes equation
\begin{equation}
\partial_{t} u + ( u , \nabla u ) - \nu \nabla^{2} u = - \frac{1}{\rho} \nabla p
\label{NavierStokes}
\end{equation}
provides the basis for the description of laminar flow. Here $u = u(x,t)$ is the velocity, $p = p(x,t)$ the pressure, and $\nu$ the kinematic viscosity. In the limit where 
$\nu \rightarrow 0^{+}$ however, the flow is known to become highly irregular, turbulence ensues, and one may ask if \eqref{NavierStokes} still captures the full dynamics. 
For such a state it is certainly necessary to consider a statistical theory where $u(x,t)$ is a random field in space and time and \eqref{NavierStokes} is replaced with an 
infinite hierarchy of dynamical equations that relate the different orders of correlation functions. Usually one has to rely on certain closure assumptions to close this set of equations 
(see e.g. \cite{MoninYaglom1975}). One may take another route however, where one starts from reasonable assumptions on the symmetries of the problem. In particular, assuming statistical 
homogeneity, isotropy, and scale-invariance, simple dimensional analysis yields a tight prediction for the moments of velocity differences (structure functions)
\begin{equation}
\overline{( \Delta_{r} u )^{n}} \propto r^{\zeta_{n}} ~, \qquad \zeta_{n} = n/3 ~,
\end{equation}
where $\Delta_{r} = \left( u(x + r) - u(x) , e_{r} \right)$, and the bar \smash{$\,\overset{\overset{\rule{15pt}{0.2pt}}{\phantom{.}}}{\,\cdots\,}\,$} denotes spatial averaging. 
The universality conjecture \cite{1941DoSSR..30..301K} then states that this scaling behavior should hold far from the boundaries and independent of the mechanisms that generate the flow.
Nevertheless, both experiment and direct numerical simulations of \eqref{NavierStokes} indicate a violation of this scaling behavior for high-order moments 
(see e.g. \cite{2002PhFl...14.1065G,2008PhRvL.100w4503B,2009AnRFM..41..165I}). In terms of symmetries this corresponds to the breaking of scale-invariance. Highly erratic, 
intermittent structures give the dominant contribution to these moments, the universal statistical properties of which are still largely unknown. What is the nature of these structures, 
and is it possible to understand their properties from first principles?

\begin{figure}[!tb]
\centering
\includegraphics[width=0.45\textwidth]{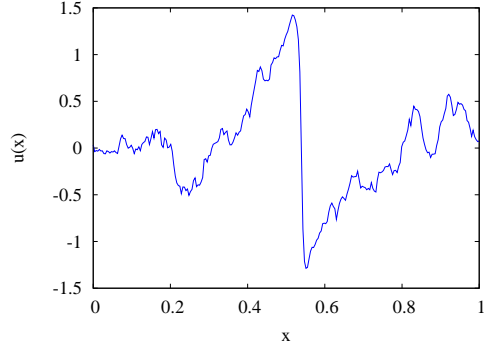}
\caption{\label{fig:Profile} Velocity profile $u(x)$ from a simulation on a \mbox{$254 \times 1024$} (space $\times$ time) lattice, where $x$ is taken in units of the spatial lattice size $L$.}
\end{figure}

Here, the random-force-driven Burgers equation  (see \cite{Burgers1973,2000nlin.....12033F,2007PhR...447....1B} for a review) is taken as a one-dimensional model
\begin{equation}
\partial_{t} u + u \partial_{x} u - \nu \partial_{x}^{2} u = f 
\label{Burgers}
\end{equation}
of the Navier-Stokes equation. We artificially generate a turbulent state by driving the system by a self-similar forcing that is white-in-time
\begin{equation}
\langle f(k,t) f(k',t') \rangle = D_{0} |k|^{\beta} \delta(k+k') \delta(t-t') ~.
\label{forcing}
\end{equation}
The brackets $\langle \cdots \rangle$ denote ensemble averaging, $D_{0}$ is a dimensionful parameter, and $\beta$ measures the relative strength of the forcing at different scales. For large,
negative values of $\beta$ the forcing effectively acts at large scales $\sim L$. On the other hand kinematic viscosity $\nu$ provides a dissipation scale $\eta$, and for 
$\nu \rightarrow 0^{+}$ these two characteristic scales separate. In particular, for $\beta = -1$ the interplay of the stochastic forcing and advective term leads to a Kolmogorov energy 
spectrum $E(k) \propto |k|^{-5/3}$ in the intermediate range of scales, reminiscent of Navier-Stokes turbulence \cite{1995PhRvE..51.2739C,1995PhRvE..52.5681C}. The physical picture that 
one may associate with this scenario is the appearance of shocks with a finite disspative width (see Fig.\,\ref{fig:Profile}). These structures give the dominant contribution to the high 
order moments of velocity differences, and leads to a strong form of intermittency 
\cite{1996PhRvE..54.4908G,1997PhRvL..78.1452B} where
\begin{equation}
\langle |\Delta u|^{n} \rangle \propto r ~, \qquad n \geq 3 ~.
\label{multifractal}
\end{equation}
In view of the well-established anomalous scaling behavior of Burgers turbulence \cite{2005PhRvL..94s4501M} and the physical picture of the underlying mechanisms for intermittency 
\cite{1995PhRvE..51.2739C,1995PhRvE..52.5681C,1996PhRvE..54.4908G,1997PhRvL..78.1452B}, 
Burgers equation \eqref{Burgers} provides an ideal benchmark setting to test new analytical and numerical methods for Navier-Stokes turbulence.

\section{Functional Integral}

The functional integral representation for the random-force-driven Burgers equation is obtained via the Martin-Siggia-Rose formalism 
by means of an auxiliary response field $\mu$ \cite{1973PhRvA...8..423M,1976ZPhyB..23..377J,1978PhRvB..18..353D,1981JSP....25..183J}. 
We have the partition function
\begin{equation}
Z = \int d [u]\, d[\mu] \exp \{ -S [u , \mu]\}~,
\end{equation}
where the action $S[u , \mu]$ is given by
\begin{equation}
S = -i \int\! dt\, dx\, \mu ( \partial_t u + u \partial_x u - \nu \partial_x^2 u )  + \, \frac{1}{2} \int\! dt\, dx\, dy \, \mu(x,t) D(x - y) \mu(y,t) ~.
\end{equation}
Here, $D(x-y)$ is the spatial part of the two-point correlation function \eqref{forcing}. Notice, that in this form the action does not satisfy positivity. To obtain a Gibbs measure that 
can be sampled by a Markov chain Monte Carlo (MCMC) algorithm we integrate out the auxiliary field. 
This leaves us with the probability density functional
\begin{equation}
P[ u ] = \exp \Big\{ - \frac{1}{2} \int\! dt\, dx\, dy \, ( \partial_t u + u \partial_x u - \nu \partial_x^2 u ) D^{-1}(x-y) ( \partial_t u + u \partial_x u - \nu \partial_x^2 u ) \Big\}~,
\label{action}
\end{equation}
which is the starting point for our investigations.

\section{Lattice Theory}

The theory is defined by placing the field $u(x,t)$ on the sites of a regular space-time lattice $\Lambda$, i.e. $(x,t) \in \Lambda$. This way, we impose a UV cutoff that 
eliminates the details of those processes occurring deep in the dissipative regime. Then, the measure is given by \mbox{$d [u] \,\rightarrow \prod_{(x,t) \,\in\, \Lambda}\! d u(x,t)$} 
and the action in \eqref{action} needs to be discretized appropriately. We replace the dynamics \eqref{Burgers} with a finite-difference equation with backward-time discretization
\begin{equation}
\partial_t u + u \partial_x u \rightarrow \frac{1}{\epsilon} (u(t) - u(t-\epsilon)) + u(t-\epsilon) \,\partial_{x} u(t-\epsilon)~,
\end{equation}
where $\epsilon$ is the lattice spacing in time direction. This ensures the correct dynamics in the continuum limit \cite{ZinnJustin:2002ru}. For the advective term we take the 
anti-symmetric spatial derivative
\begin{equation}
\partial_{x} u \rightarrow \frac{1}{2 a} ( u(x+a) - u(x-a))~,
\end{equation}
where $a$ is the lattice spacing in the spatial direction. With this choice of discretization the problem is amenable to a local over-relaxation algorithm \cite{2008EL.....8440002D}. 
Starting from an initial configuration $\{ u(x,t) , (x,t) \in \Lambda \}$ the set of single-site variables is updated iteratively by the successive application of a 
transition probability $P(u(x,t) \rightarrow u'(x,t))$. We use the high-quality \verb|ranlux| (pseudo) random number generator \cite{1994CoPhC..79..100L} which is essential 
for large-scale lattice simulations. Specific improvements, e.g. Chebyshev acceleration \cite{1962mia..book.....V} significantly reduce thermalization and autocorrelation
times for the relevant observables.

In our simulations we use periodic boundary conditions in space and fixed (Dirichlet) boundary conditions in time. That way we eliminate the zero mode from the dynamics. One important point
is that the probability distribution functional \eqref{action} defines a stationary process for a system of infinite extent (in the time direction), i.e.
\begin{equation}
\langle u(x_1, t) u(x_{2},t) \cdots u(x_{n},t) \rangle = \langle u(x_{1}, t+t') u(x_{2} , t + t') \cdots u(x_{n} , t+t' ) \rangle ~.
\end{equation}
In practice, this condition has to be checked explicitly. We find that for a finite space-time lattice this property holds to good approximation in the middle of the configurations where 
boundary effects are neglible. This defines the physical region where one may extract correlation functions.

Another issue is that of Galilean invariance \cite{2007PhRvL..99y4501B,2009NuPhB.814..522B}. Both the action \eqref{action} and the measure are invariant in the continuum under Galilean 
transformations
\begin{equation}
x \rightarrow x + r ~, \quad u(x) \rightarrow u(x+r) + v ~, \quad r = v \, t ~.
\label{GT}
\end{equation}
To avoid an overcounting of field configurations one has to perform a gauge fixing, where one inserts
\begin{equation}
1 = \mathcal{J}[u] \int dv \, \delta[u(r = v \, t , t) + v]  ~.
\end{equation}
in the dynamic functional \eqref{action}. Here, $\mathcal{J}[u]$ is the Faddeev-Popov Jacobian. While gauge fixing is unavoidable for generic correlation functions this is not so for velocity 
differences that are clearly invariant under \eqref{GT}.

We want to give a short remark on the computational requirements. Since we use a local over-relaxation algorithm, the long-range correlations imposed by the 
forcing \eqref{forcing} prohibit any attempt to parallelize in the spatial direction. This poses a severe problem when turning to higher dimensions and it is
absolutely necessary to switch to a global, e.g. Hybrid Monte Carlo algorithm. With a parallel code (in the time direction) for a $245 \times 1024$ lattice (space $\times$ time) our 
simulations currently run on up to 512 processors.

\section{Results}

\begin{figure*}[!t]
\subfigure{\includegraphics[width=0.45\textwidth]{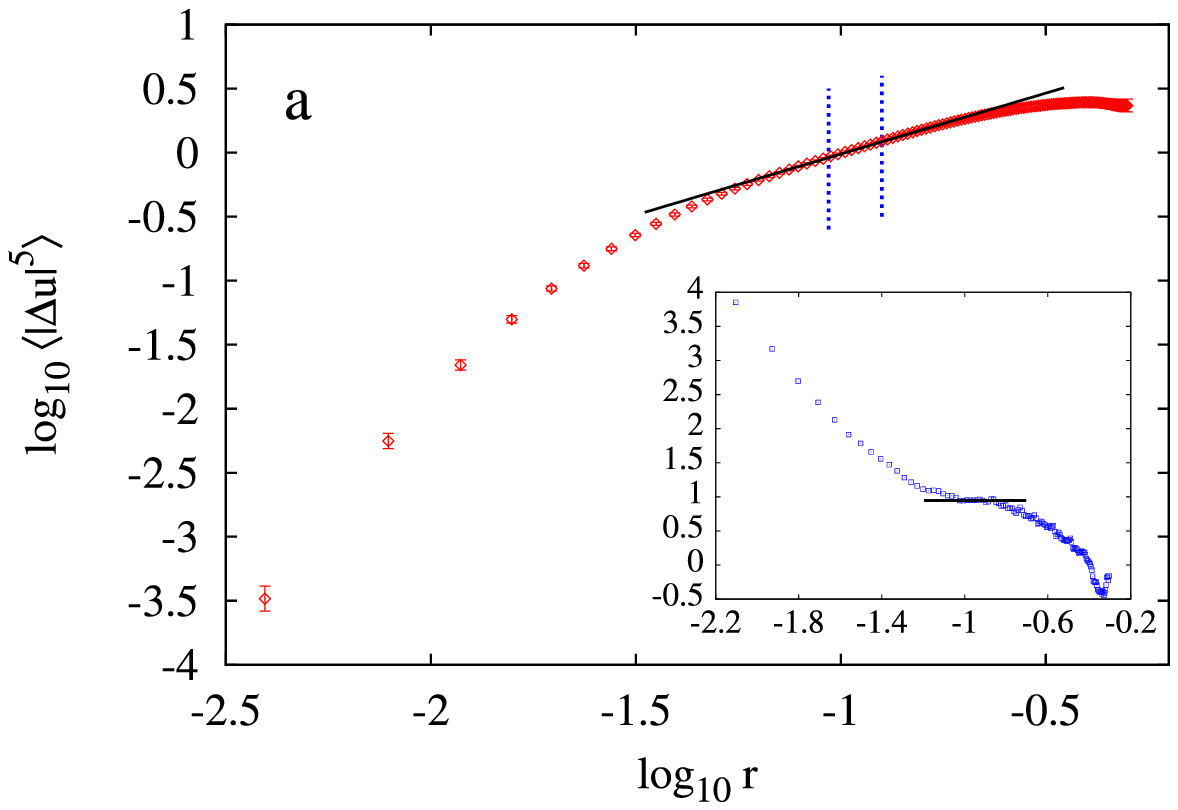}}
\subfigure{\includegraphics[width=0.45\textwidth]{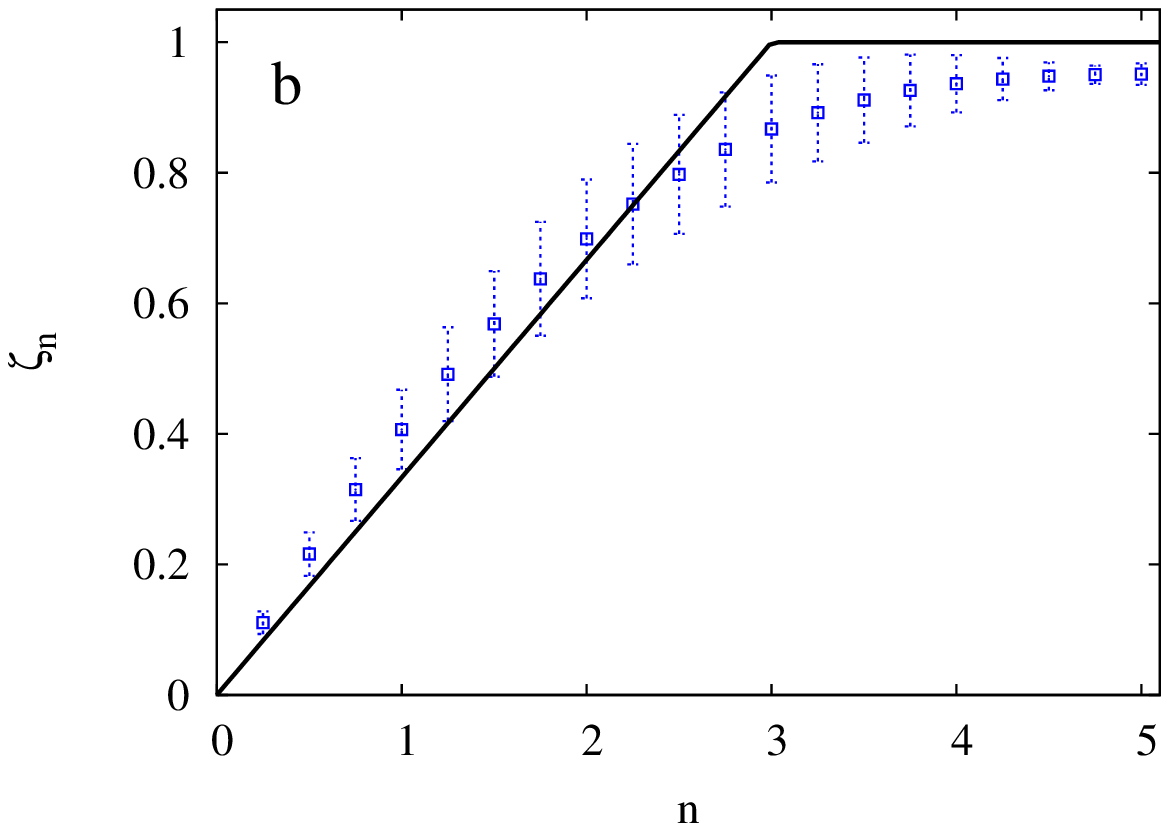}}
\caption{\label{fig:StructureFunction} (a) Log-log plot of the structure function of order $n = 5$ with a linear scaling function plotted for comparison. 
Vertical bars indicate the region for the extraction of scaling exponents. Inset shows the local slopes versus $r$. 
(b) Structure function scaling exponents $\zeta_n$ versus order $n$. The black curve indicates a bifractal scaling behavior.}
\end{figure*}

Structure functions are evaluated over an ensemble of configurations generated by the MCMC algorithm. We measure structure functions in the middle of our configurations at randomly 
chosen starting points. That way it is possible to reduce autocorrelation effects significantly. The main results of our simulations are shown in Fig.\,\ref{fig:StructureFunction}a and 
Fig.\,\ref{fig:StructureFunction}b. For details on the extraction of structure functions and the scaling spectrum we refer to \cite{2011NJPh...13j3028M}.
Here, as an example we show a log-log plot of the fifth order structure function (see Fig.\,\ref{fig:StructureFunction}a). The scaling region is clearly visible, and we have indicated the 
region for the extraction of the scaling exponents by two vertical lines. In practice, we are bound to work at finite viscosity, and at small values of the separation we see the dissipative 
regime where the scaling breaks down. For comparison, in the inset we have plotted the local scaling exponents evaluated over three successive points. 
One may recognize, that in the scaling region the values lie on a plateau (as indicated by the horizontal line) which defines the scaling exponent. Applying this procedure to all
structure functions of order $n < 5$ yields the scaling spectrum shown in Fig.\,\ref{fig:StructureFunction}b. The black line shows the bifractal scaling prediction \eqref{multifractal}.
Our results are in good agreement with this prediction and also with previous results from high-resolution simulations employing a fast Legendre transform algorithm 
\cite{2005PhRvL..94s4501M}.

\begin{figure*}[!t]
\centering
\subfigure{\includegraphics[width=0.45\textwidth]{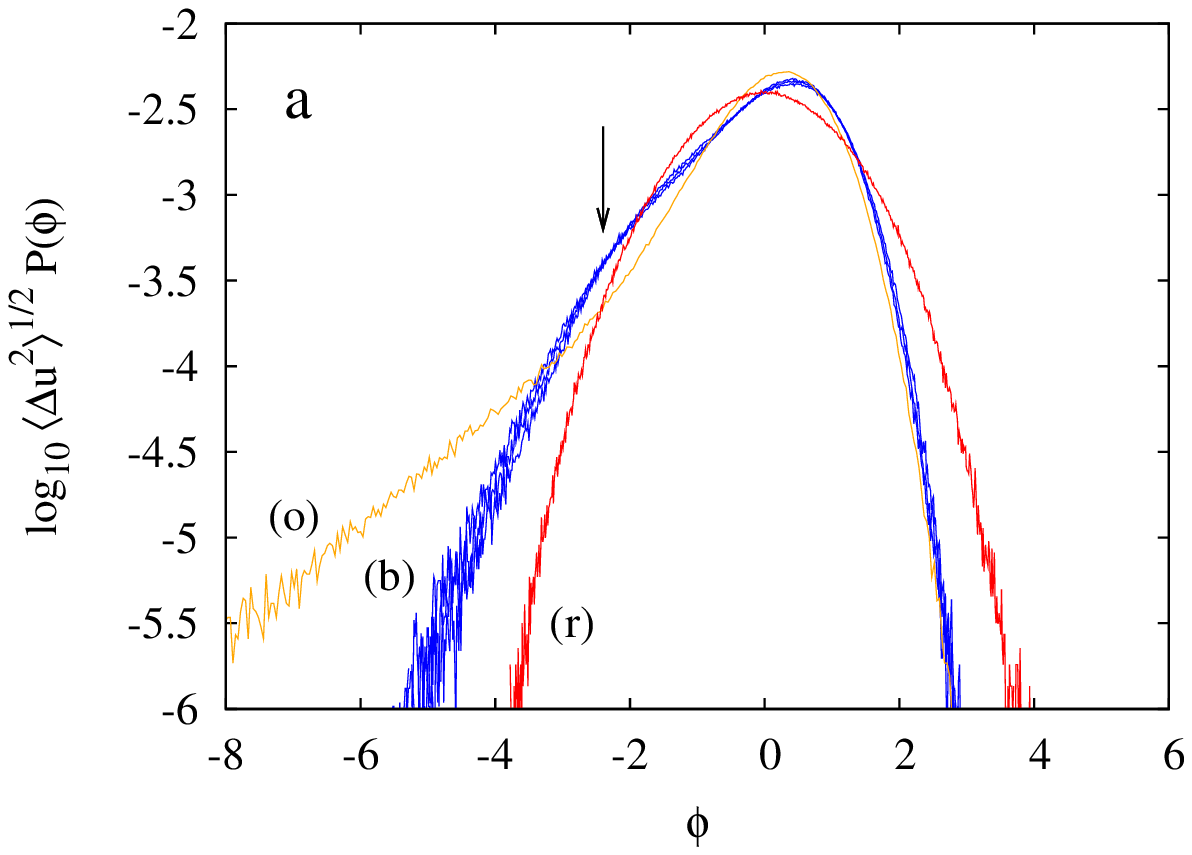}}\qquad
\subfigure{\includegraphics[width=0.45\textwidth]{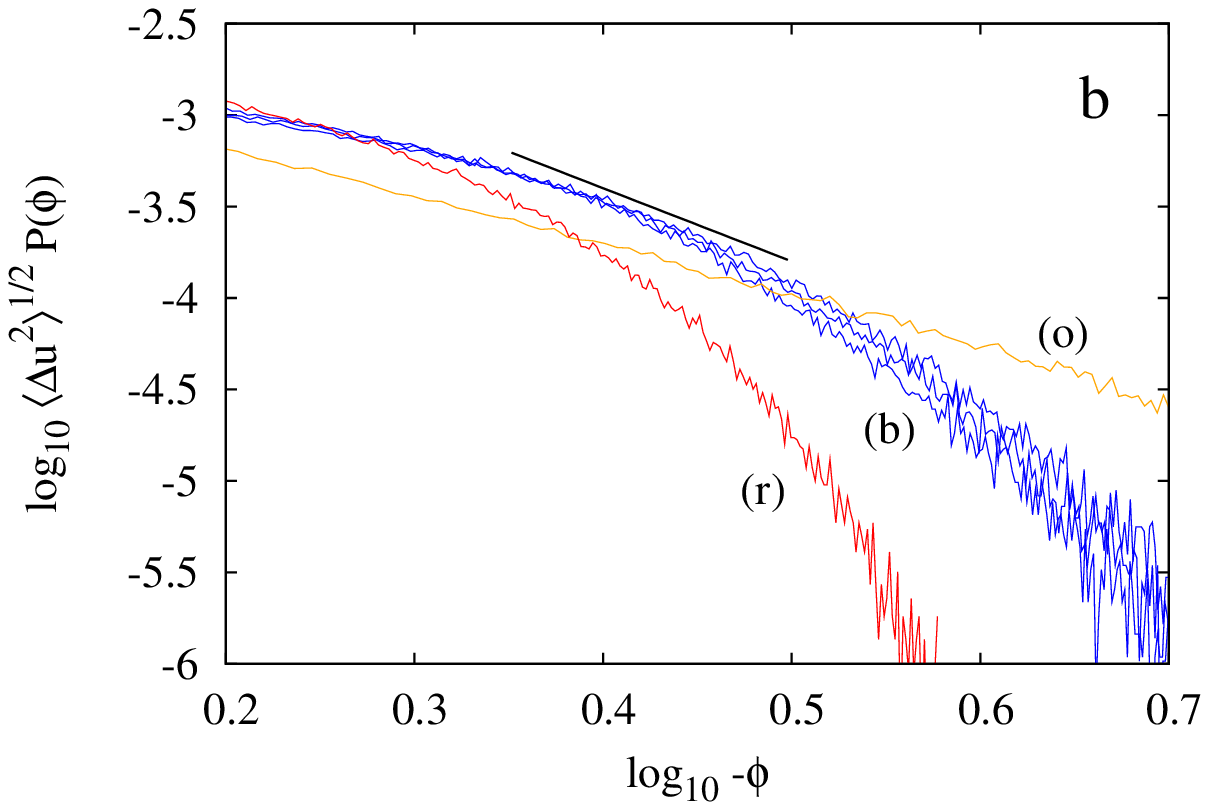}}
\caption{\label{fig:PDF1} Probability distribution functions $\mathcal{P}(\Delta u , r)$ as a function of the dimensionless variable
$\phi = \Delta u / [ \langle \Delta u^{2} \rangle ]^{1/2}$ plotted for different values of $r$. (a) Collapse of the PDF in the universal regime (blue). 
In the energy-containing range (red) the fluctuations become Gaussian -- the random forcing dominates -- whereas in the dissipative regime (orange) 
fluctuations are strongly enhanced. (b) Scaling region for the left tail of the PDF. The black line indicates the scaling prediction with exponent $\gamma = -4$.}
\end{figure*}

We can extract important information on the physical behavior from the probability distribution functions (PDF) of velocity differences $\mathcal{P}(\Delta u , r)$. In Fig.\,\ref{fig:PDF1}a we 
show the PDF of velocity differences $\Delta u = u(x + r) - u(x)$ for different values of the separation $r$ plotted as a function of the dimensionless variable
$\phi = \Delta u / [ \langle \Delta u^{2} \rangle ]^{1/2}$. One may clearly recognize the influence of the random forcing acting at large scales (red) where the fluctuations become Gaussian. 
For smaller values of the separation large fluctuations become strongly enhanced by the dynamics (blue, orange). In particular, in the disspative regime (orange), at very small separations,
these fluctuations are directly associated with the dissipative shocks (see Fig.\,\ref{fig:Profile}). In the intermediate range the PDF collapse (blue curves) and we have an indication of universal behavior.

One particularly interesting region is indicated by the arrow in Fig.\,\ref{fig:PDF1}a. Here, the PDF for different values of the separation collapse exactly -- this corresponds to the region 
\mbox{$|\Delta u| \ll u_{rms}$}, $r \ll L$ where the PDF of velocity differences has the universal scaling form
\begin{equation}
\mathcal{P} (\Delta u , r) = r^{-z} f\left( \Delta u / r^{z} \right)
\end{equation}
with the dynamic exponent $z$. In the asymptotic region $- \Delta u / r^{z} \gg 1$ where $\Delta u < 0$ we expect the algebraic
scaling
\begin{equation}
\mathcal{P} (\Delta u ,r) \propto \left(\Delta u \right)^{\gamma} ~,
\end{equation}
with $\gamma = -4$ \cite{PhysRevE.52.6183,1996PhRvL..77.3118Y}. This is shown in Fig.\,\ref{fig:PDF1}b where we have plotted the scaling region of the left tail of the PDFs. Though our data is not sufficient to
clearly extract the scaling exponent, our results are in agreement with the scaling prediction.

\section{Summary}

We have demonstrated that lattice simulations can contribute to the understanding of intermittency in turbulence. Our simulations clearly show anomalous scaling for the high order
moments of velocity differences where the exponents are in excellent agreement with previous estimates \cite{2005PhRvL..94s4501M}. We want to emphasize that in terms of computational efficiency our
method cannot compete with other conventional time-advancing methods, e.g. pseudo-spectral or finite-difference methods. However, lattice simulations may provide a different perspective on
the problem of intermittency where large fluctuations play a dominant role \cite{1996PhRvE..54.4908G,1997PhRvL..78.1452B}.

\end{document}